\documentstyle[preprint,tighten,aps]{revtex}

\preprint{\vbox{\baselineskip=12pt
\rightline{CGPG-98/3-3}}}

\begin{document}
\draft
\title{Quanta of Geometry and Rotating Black Holes}
\author{K.\ Krasnov\thanks{E-mail address: krasnov@phys.psu.edu}}
\address{Center for Gravitational Physics and Geometry, \\
The Pennsylvania State University, PA 16802\\ and \\
Institute for Theoretical Physics,
University of California, Santa Barbara, CA 93106.}

\maketitle

\begin{abstract}

In the loop approach to quantum gravity the spectra of operators
corresponding to such geometrical quantities as length, area
and volume become quantized. However, the size of arising
quanta of geometry in Planck units is not fixed by the
theory itself: a free parameter, sometimes referred to as
Immirzi parameter, is known to affect the spectrum of
all geometrical operators. In this paper I propose an
argument that fixes the value of this parameter. I consider 
rotating black holes, in particular the extremal ones.
For such black holes the ``no naked singularity
condition'' bounds the total angular momentum $J$ by
$A_H/8\pi G$, where $A_H$ is the horizon area and $G$ Newton's
constant. A similar bound on $J$ comes from the quantum theory.
The requirement that these two bounds are the same
fixes the value of Immirzi parameter to be unity. A byproduct
of this argument is the picture of the quantum extremal rotating
black hole in which all the spin entering the extremal
hole is concentrated in a single puncture.

\end{abstract}
\pacs{}

Even a most naive application of quantum ideas to gravity suggests that
the picture of continuous fabric of spacetime becomes meaningless on 
distances of the order of Planck length. Current 
approaches to quantum gravity support this conclusion.
Thus, for example, in string theory, the usual 
description of geometry in terms of spacetime metric whose evolution is
governed by some action principle is valid only at low energies.
An old set of results \cite{Strings} on high energy string scattering 
amplitudes indicates that at very high energies the size of strings 
becomes important which means that the usual metric description of spacetime 
geometry is not good at short distances. In non-perturbative
quantum gravity, or loop quantum gravity, as it is sometimes called,
the discrete structure of geometry on Planck scale is manifested
by the fact that spectra of operators corresponding to such
geometrical quantities as length, area and volume become quantized 
\cite{Quantized}. The ``size'' of arising quanta of geometry is
proportional to Planck length. However, the spectra of geometrical
operators turn out to depend on an additional physical parameter,
not fixed by the theory itself, which determines the size of 
quanta of geometry in Planck units. The presence of this parameter
in the theory was pointed out by Immirzi \cite{Imm}, and I shall
refer to it as Immirzi parameter.

In this paper I argue that the value of Immirzi parameter in 
loop quantum gravity must be set to a particular fixed value.
This completely fixes the size of quanta of geometry 
in the approach. I make the argument by comparing some results
from loop quantum gravity with some facts about
rotating black holes, in particular the extremal rotating black 
holes. As a byproduct of this argument we shall arrive at 
an interesting quantum description of the extremal 
rotating black holes.

In loop quantum gravity, the elementary excitations of geometry are
described in terms of one-dimensional objects - loops, or, more
generally, graphs in space. The edges of these graphs carry
various quantum numbers, in particular spin, and can be
thought of as flux lines. In particular, these lines ``carry''
area: surfaces
acquire area through their intersections with the flux lines.
The flux of area carried by each flux line is quantized,
which implies the quantization of the area spectrum in the
theory. The spectrum is given by \cite{Quantized}
\begin{equation} \label{area}
A_S = 8\pi\gamma l_p^2 \sum_p \sqrt{j_p(j_p+1)}.
\end{equation}
Here $A_S$ stands for an eigenvalue of the operator measuring
the area of a surface $S$, sum is taken over all points $p$
where flux lines intersect the surface, $j_p$ are
spins (half-integers) labelling the corresponding flux lines,
$\gamma$ is a real, positive parameter (Immirzi parameter)
and $l_p^2=G\hbar$, $G$ being Newton constant. We set $c=1$
throughout. The spectrum is purely discrete, and $\gamma$
fixes the size of quanta of area in Planck units. Let us 
emphasize that $l_p$ in (\ref{area}) is the usual Planck
length $l_p=\sqrt{G\hbar}\approx1.6\times10^{-33}{\rm cm}$ 
calculated using  the physical, ``macroscopical'' 
Newton's constant. Due to renormalization effects, the later 
may, in principle, be different from the ``microscopical'' Newton's 
constant that enters as a parameter of the quantum theory.
If such a renormalization occurs, the corresponding
renormalization factor is absorbed in $\gamma$ so that
$l_p$ is the usual Planck length.

To fix the value of parameter $\gamma$ let us consider the
Kerr family of black hole solutions. Let us recall some
simple facts about these black holes. They are described
by two parameters: mass $M$ and angular momentum $J$. The
horizon area $A_H$ is given by the well-known function of $M,J$.
The condition that there is no naked singularity reads
$J/M \leq GM$. The value of $J$ saturating the inequality
corresponds to the extremal black hole. Instead of using 
$M,J$ as the independent parameters describing the black hole, 
one can work with $A_H, J$.
A simple algebraic manipulation shows that
``no naked singularity'' condition expressed in 
terms of the parameters $A_H, J$ reads
\begin{equation} \label{cond}
J \leq {A_H\over 8\pi G}.
\end{equation}
The value of angular momentum $J$ saturating the 
inequality corresponds to the extremal black hole.

The argument I am about to present depends on two
assumptions. The first one is that the area spectrum
for the horizon surface of a rotating black hole is given
by (\ref{area}). This can almost be said to be
a prediction of loop quantum gravity, because
the spectrum (\ref{area}) is derived from very
general principles and should be valid for any
surface, in particular the horizon. On the
other hand, we do not have a quantum description
of rotating black holes as of yet, so the validity
of (\ref{area}) in this context is still an assumption.
The second assumption is much stronger. I assume
that the total angular momentum of the black hole is
related to the spins labelling the flux lines intersecting
the horizon. It is then most natural to assume that
the black hole angular momentum is given by a vector
sum of spins entering the hole. A precise way the
angular momentum is obtained from spins may be very
complicated. However, independently of these details,
the total angular momentum will satisfy the inequality
\begin{equation}\label{ass}
J \leq \hbar\sum_p j_p.
\end{equation}
The argument below depends crucially on the assumption expressed
by this inequality.

Let us now consider a family of quantum black holes for
which the algebraic sum of spins $\sum_p j_p$ is the same.
According to (\ref{ass}), the angular momentum for
all these black holes is bound by $\hbar\sum_p j_p$.
On the other hand, the area spectrum (\ref{area}) tells
us that the horizon area $A_H>8\pi\gamma l_p^2 \sum_p j_p$.
Thus, what we get is a bound on possible values of 
angular momentum of the black hole of a fixed horizon
area:
\begin{equation}\label{in}
J < {A_H\over 8\pi\gamma G}.
\end{equation} 
The above inequality is strict, that is one can never
saturate the bound. However, there is a way to
arrange punctures such that the bound is ``almost''
saturated. Indeed, by putting all the spin entering
the black hole in a single puncture of spin $j$ we get 
$A_H = 8\pi\gamma l_p^2 j + O(1)$. Thus, for a large 
$j$ the bound in (\ref{in}) can be ``almost'' saturated
by putting all the spin at a single puncture.

The bound (\ref{in}) is strikingly resemblant of the
inequality (\ref{cond}). The values of parameters
in (\ref{cond}) that saturate the bound correspond to the
extremal black holes. Quite similarly, the quantum
states that are closest to the saturation of the
bound in (\ref{in}) are the ``extremal'' ones
for which all the spin entering the hole is 
concentrated in a single puncture. This strongly
suggests that the two bounds --the one coming
from the properties of the classical black hole
solution and the one coming from the quantum
theory-- express the same physical property and
must coincide. This is only possible if one fixes
the value of the Immirzi parameter to be the unity:
\begin{equation}\label{one}
\gamma=1.
\end{equation}
This is the end of the argument.

Let us conclude with several remarks. (i) The value
(\ref{one}) of the Immirzi parameter is {\it not}
the same value as one gets from the quantum mechanical
calculation of black hole entropy \cite{ABCK}.
In fact, with the above value of the parameter,
the approach of \cite{ABCK} gives the value for
the entropy $S\approx 0.2 S_{BH}$, where $S_{BH}$
is Bekenstein-Hawking entropy. Thus, either the
argument presented has a flaw, or the approach
\cite{ABCK} undercounts the states. (ii) The
above argument depends crucially on the assumption
that the angular momentum of a rotation black hole
is related to the spin entering the hole. It is not
at all obvious that this must be the case, because
the angular momentum of the hole is related to 
spacetime rotations, while the spin labelling the
flux lines has to do with rotations in the 
internal space. On the other hand, there are no 
other quantities to which the angular momentum can be
related: in the absence of matter the spins (and the
related labels for the intertwiners) are
the only quantum numbers labelling the states. 
Thus, whether the above assumption is true or false
remains to be seen. (iii) It is encouraging that
our assumption leads to such a simple picture
for extremal black holes. As we saw, the extremal 
rotating black hole is described by the ``extremal''
configuration in which all the spin entering the
hole is concentrated at a single point on the horizon.
At the same time all the area is concentrated at
the same point. Thus, the horizon geometry becomes
highly non-classical at extremality.

\bigskip
\bigskip
{\bf Acknowledgements:} I am grateful to T. Jacobson for 
a discussion and to L. Freidel for reading the manuscript.
This work was supported in part by 
the Braddock fellowship of Penn State, by the NSF
grants PHY95-14240, PHY94-07194 and by the Eberly 
research funds of Penn State. 
The author is also grateful to the Institute for Theoretical
Physics, Santa Barbara, where this work was completed.


\begin{thebibliography}{99}
\bibitem{Strings} D.\ Amati, M.\ Ciafaloni, G.\ Veneziano,
Can spacetime be probed below the string size?,
{\sl Phys.\ Lett.} {\bf B216} 41 (1989).

D.\ J.\ Gross, P.\ F.\ Mende, String theory beyond the Planck scale, 
{\sl Nucl.\ Phys.} {\bf B303} 407 (1988).

\bibitem{Quantized} C.\ Rovelli and L.\ Smolin, Discreteness of area 
and volume in quantum gravity, {\sl Nucl.\ Phys.} {\bf B442},
593 (1995); Erratum: {\sl Nucl.\ Phys.} {\bf B456}, 734 (1995).

R.\ De Pietri and C.\ Rovelli, Geometry Eigenvalues and Scalar Product 
from Recoupling Theory in Loop Quantum Gravity, {\sl Phys.\ Rev.} {\bf D54},
2664 (1996).

T.\ Thiemann, A length operator for canonical 
quantum gravity, available as gr-qc/9606092.

A.\ Ashtekar and J.\ Lewandowski, Quantum Theory of Geometry I: Area
operators, {\sl Class.\ Quant.\ Grav.} {\bf 14}, 55 (1997). 

A.\ Ashtekar and J.\ Lewandowski, Quantum Theory of Geometry II:
Volume Operators, {\sl Adv.\ Theor.\ Math.\ Phys.} {\bf 1}, No. 2 (1998).

\bibitem{Imm} G.\ Immirzi, Quantum Gravity and Regge Calculus,
{\sl Nucl.\ Phys.\ Proc.\ Suppl.} {\bf 57} 65-72 (1997).

\bibitem{ABCK}  A.\ Ashtekar, J.\ Baez, A.\ Corichi, K.\ Krasnov, 
Quantum Geometry and Black Hole Entropy,
{\sl Phys.\ Rev.\ Lett.} {\bf 80}, No. 5, 904-907 (1998);

A.\ Ashtekar and K.\ Krasnov, Quantum Geometry and Black Holes,
in BLACK HOLES, GRAVITATIONAL RADIATION AND THE UNIVERSE,
Essays in honor of C.V. Vishveshwara, Ed. B. R. Iyer and B. Bhawal, 
Kluwer, Netherlands;
also available as gr-qc/9804039.

\end{thebibliography}
\end{document}